\documentclass[journal]{IEEEtran}

\usepackage{graphicx} 
\usepackage{algorithm}
\usepackage{algorithmic}
\usepackage{multirow}
\usepackage{threeparttable}
\usepackage{booktabs}
\usepackage{tabularx} 
\usepackage{colortbl}  
\usepackage{bm}
\usepackage{amsmath}
\usepackage{amssymb}
\usepackage{cite}
\usepackage{hyperref}
\usepackage{xcolor}
\usepackage{url}
\ifCLASSINFOpdf
\else
\fi
\begin{document}
%
\title{Audio Tagging by Cross Filtering Noisy Labels}
%
%
%

\author{Boqing~Zhu, Kele~Xu, Qiuqiang~Kong, Huaimin~Wang and~Yuxing~Peng

\thanks{Boqing Zhu, Kele Xu, Huaimin Wang and Yuxing Peng are with the Science and Technology on Parallel and Distributed Laboratory, National University of Defense Technology, Changsha, 410073, China. (e-mails: zhuboq@gmail.com, kelele.xu@gmail.com)

QiuQiang Kong is with the Centre for Vision, Speech and Signal Processing, University of Surrey, Guildford GU2 7XH, UK.

(\emph{Corresponding Author: Kele~Xu.})
}


}

\maketitle

\begin{abstract}

High quality labeled datasets have allowed deep learning to achieve impressive results on many sound analysis tasks. Yet, it is labor-intensive to accurately annotate large amount of audio data, and the dataset may contain noisy labels in the practical settings. Meanwhile, the deep neural networks are susceptive to those incorrect labeled data because of their outstanding memorization ability.
In this paper, we present a novel framework, named CrossFilter, to combat the noisy labels problem for audio tagging.
Multiple representations (such as, Logmel and MFCC) are used as the input of our framework for providing more complementary information of the audio. 
Then, though the cooperation and interaction of two neural networks, we divide the dataset into curated and noisy subsets by incrementally pick out the possibly correctly labeled data from the noisy data.
Moreover, our approach leverages the multi-task learning on curated and noisy subsets with different loss function to fully utilize the entire dataset. The noisy-robust loss function is employed to alleviate the adverse effects of incorrect labels.
On both the audio tagging datasets FSDKaggle2018 and FSDKaggle2019, empirical results demonstrate the performance improvement compared with other competing approaches. On FSDKaggle2018 dataset, our method achieves state-of-the-art performance and even surpasses the ensemble models.

\end{abstract}

\begin{IEEEkeywords}
Audio tagging, noisy labels, deep convolutional neural network, cross representation, DCASE challenge.
\end{IEEEkeywords}

%
\IEEEpeerreviewmaketitle

\section{Introduction}
\label{sec:intro}
%
%
%
%

\IEEEPARstart {A}{udio} tagging aims to identify the presence of sound events in an audio recording. For different sound analysis tasks, audio tagging has drawn lots of attentions as its applications seems to be evident in many different fields, such as multimedia indexing and retrieval \cite{kiranyaz2006generic}, surveillance and monitoring application \cite{crocco2016audio}. Since the revolution of neural networks, supervised learning with deep neural networks (DNNs) has become a common approach for the classification task. The success of DNNs is highly tied to the large and carefully annotated dataset. Unfortunately, manually labeling large-scale dataset is expensive and time-consuming for the auditory data. Several inexpensive ways can be employed to collect labeled data, including online queries and crowdsourcing. Those approaches invariably yield a large number of noisy (incorrect) labels. Moreover, even manually labeled data are likely to have incorrect labels as data labeling is subjective and error-prone. Noisy labels which corrupt from the ground-truth labels are ubiquitous in the piratical settings. For instance, AudioSet \cite{gemmeke2017audio} consists of 5000 hours labeled with 527 classes, but label error is estimated at above 50\% for about 18\% of the classes. As a trade-off between manual verification cost and the size of dataset, many current approaches use a bi-quality dataset for training in the practical settings. Here, bi-quality dataset means only a small portion of data is accurately labeled, while the rest is corrupted (contain noisy labels).

Sustainable efforts have been made for the learning with noisy labels (LNL) problem. Specifically, a number of approaches focus on estimating the noise transition matrix. However, the noise transition matrix is hard to be estimated precisely when the number of classes becoming large (the number of sound events is naturally large). Without estimating the noise transition matrix, ``co-learning'' is one promising learning paradigm to combat with the noisy labels. The primary idea of ``co-learning'' is that two neural networks can robustly learn from noisy labeled data by cooperating and interacting with each other. Co-teaching \cite{han2018co} and WeblyNet \cite{kumar2018learning} are two well-known approaches inspired by Co-training \cite{blum1998combining}. An important character for these approaches is to keep two peer networks in the system ``good and different''. The diversity of the two networks in Co-teaching mainly comes from (random) weights initialization. However, they are easy to converge to consensus and lead to the inability for selecting correctly labeled data in the training process. The different ``views'' in WeblyNet are two kinds of bottleneck features of same input audio representation. The dissimilarity between them is insufficient (more comparison in Section \ref{sub:relations_to_previous_works}). To avoid accumulated error caused by sample-selection bias, increasing the diversity between the peer networks is important for these approaches.

Learning with noisy labels has attracted many research interests in the computer vision filed \cite{natarajan2013learning,wang2018iterative,zhang2018generalized,veit2017learning}. Still, audio tagging under noisy labels is an under-explored problem. In this paper, we present a novel LNL framework for audio tagging, named CrossFilter. Similar to previous ``co-learning'' approaches, we also maintain two network branches simultaneously. Our contributions are threefold:
Firstly, to increase the diversity between two peer networks, multiple audio representations are used to describe the divergence characteristics. Although there are various representations of audio, it is unclear which representation and combination is better choice for audio tagging tasks. We evaluate the commonly used representations including: linear-frequency power spectrogram (Spec), Mel-frequency cepstral coefficients (MFCC), Log-scaled Mel-spectrograms (Logmel) and constant Q spectral transform (CQT).
Secondly, the Noise Filtering method is proposed to incrementally pick out the possibly correctly labeled data from the noisy data for each peer network. It captures the simple intuition: if the prediction of an audio clip is consistent with the given label, we think the label is more credible and pick it out. Then, it will be used to train another network. Since two networks have learned different representations, they can filter different corruption introduced by noisy labels. Moreover, the sample selection is done even without knowing the ratio of noisy labels and in a class-balanced way. 
Thirdly, we explore multi-task learning (MTL) to improve the generalization performance. There are two motivations here: the first motivation is that the distributions of the correctly labeled data and noisy data are mismatch. The classifier which mapping from the feature space to tags would not fit the curated data well due to the influence of noisy labels. For another motivation, Co-teaching and other selective sampling approaches \cite{malach2017decoupling,yu2019does} only employ the selected data for training. As it is hard to accurately select out all the correctly labeled data, the discard of noisy data may reduce the number of valuable samples. By employing MTL, the network could fully utilize both correctly labeled data and the noisy data and attenuate the adverse effects of incorrect instances.

To demonstrate the efficacy of the proposed method, we conduct extensive experiments on two widely-used bi-quality audio tagging datasets. On FSDKaggle2018 dataset, we obtain an mAP@3 of 95.59\%, which achieves the state-of-the-art performance and even surpasses the ensemble models. We achieve a lwlrap score (defined in Section \ref{sub:metrics}) of 0.7195 on FSDKaggle2019 dataset, which is competitive to ensemble-based methods.

The rest of this paper is organized as follows. We first introduce related work in the following Section \ref{sec:related_work}, then present the details of the proposed approach in Section \ref{sec:method}. The experimental procedures and results in given in the Section \ref{sec:Experiments} and Section \ref{sec:Results}. The discussion and conclusion are given in Section \ref{sec:conclusion}.

\section{Related Works}
\label{sec:related_work}

\begin{figure}[t]
\centering
\includegraphics[width=0.95\columnwidth]{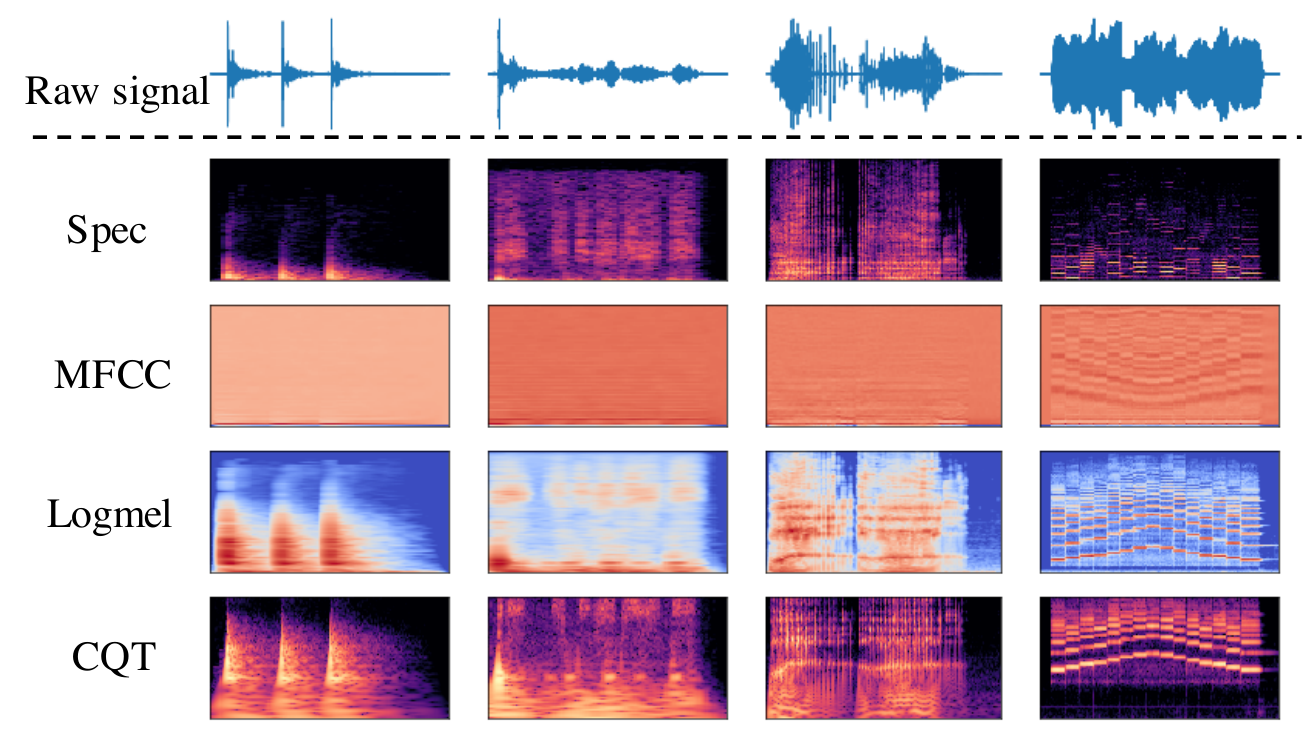} 
\caption{Different representations of audio signals: Raw wave signal, linear-frequency power spectrogram (Spec), Mel-frequency cepstral coefficients (MFCC), Log-scaled Mel-spectrograms (Logmel) and constant Q spectral transform (CQT).}
\label{fig1}
\end{figure}
\subsection{Learning with Noisy Labels}

Noise corruption is a common phenomenon within the large-scale dataset. The systematic error introduced by the label corruption is not negligible when the percentage of noisy labeled data increases. To combat with noisy labels, several methods have been proposed. Generally speaking, they can be categorized into four kinds: loss correction, transition matrix estimation, consistency regularization and selective sampling-based method.

\textbf{Loss correction}. For the loss correction strategy, the robustness of models against noisy label can be enhanced by adjusting the loss function. For example, a method \cite{reed2014training} is proposed to change the cross entropy loss function by adding a regularization term which takes the current prediction into account. 
The batch-wise loss masking is proposed in \cite{jeong2018audio}, it ignores the large losses when upgrading weights. 
The noisy-robust loss function \cite{fonseca2019learning} is firstly applied to the sound event classification task. 
Nonetheless, these surrogate losses have difficulty in optimizing the DNNs for high-dimension data.

\textbf{Transition matrix estimation}. 
This kind of method assumes that noisy labels are corrupted from the ground-truth by an unknown noise transition matrix. The matrix provides the probability of each class being mislabeled into another. By accurately estimating this matrix, the accuracy of classifiers can be improved.
For example, a two-step solution \cite{patrini2017making} to estimating the noise transition matrix heuristically and a human-assisted approach \cite{han2018masking} that conveys human cognition of invalid class transitions. However, the noise transition matrix is hard to estimate accurately when the number of classes becoming large.

\textbf{Consistency regularization}. In consistency regularization category, self-ensembling \cite{laine2016temporal} can generate a consensus prediction of the unknown labels using the outputs of the network. Instead of averaging the label predictions, Mean Teacher \cite{tarvainen2017mean} averaged model weights which enforced the smoothness of the model parameters. Yet, this kind of methods require extra computational resources to train multiple models in parallel.

\textbf{Selective sampling}. Selective sampling-based methods aim to reduce noisy ratio by selecting samples. MentorNet \cite{jiang2017mentornet} proposed to learn the curriculum from data by another neural network, which deployed a mentor net to select samples for training with noisy labels. Co-teaching+ \cite{yu2019does} suggested to improve the Co-teaching by adding the ``update by disagreement'' strategy. This approach requires to know the ratio of noisy labels, which is not always available in practical settings. Our proposed method is a form of selective sampling-based methods.


\begin{figure*}[t!]
\centering
\includegraphics[width=1.95\columnwidth]{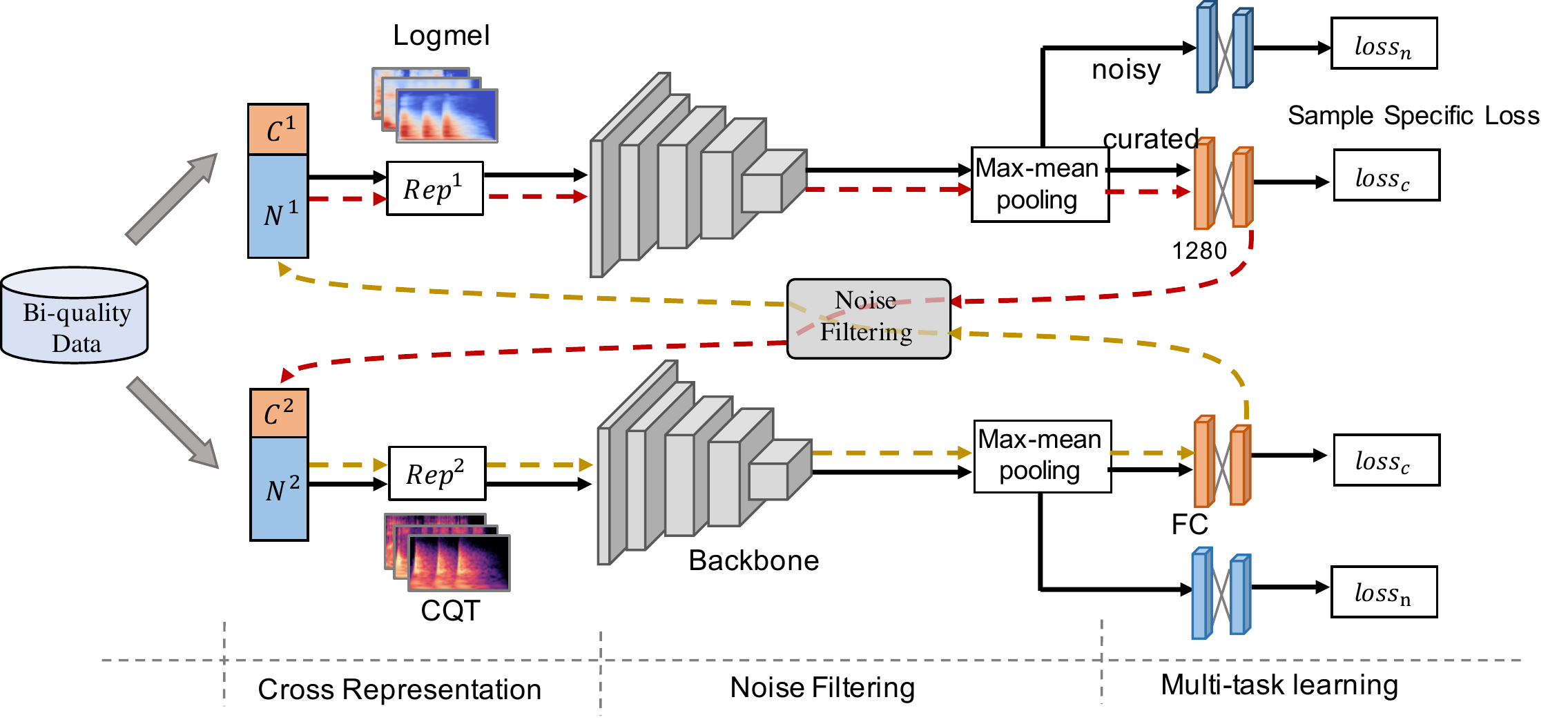} 
\caption{\textbf{Overview of CrossFilter}. Firstly, each network feeds forward a kind of representation of all data and picks out possibly correct labels (dashed lines). Secondly, remove the possibly correct labeled data from noisy subset $\mathcal{N}^r$ and add them to the curated subset $\mathcal{C}^r$ for peer network. Thirdly, each network trains with the disjoint subsets yield by its peer network (solid line). The networks are jointly optimized by two types of losses: $loss_c$ for clean data and $loss_n$ for noisy data.}
\label{fig:overview}
\end{figure*}

\subsection{Audio Tagging under Noisy Labels}

Many methods have been proposed on audio tagging, a review in this field is present in \cite{virtanen2018computational} and the recent developments in deep learning for audio signal processing is introduced in \cite{hendrik2019Deep}. Most of the previous works \cite{xu2017unsupervised,yin2019multi,xu2018mixup,zhu2018learning,Akiyama2019} mainly transform a raw signal to one single representation. 
For example, Figure \ref{fig1} shows some widely used representations including the Spec, MFCC, Logmel and CQT. Then the DNN-based classifiers can be trained on one of them. Multiple representations are applied to build an audio tagging system by ensembling convolutional neural networks (CNNs)\cite{wilkinghoff2018general,weiping2017}. Nevertheless, how to fully employ the complementary information from multiple representations is still under-explored. 

Audio Tagging with noisy label is included in the DCASE Challenges \cite{Fonseca2018_DCASE} from 2018. The tasks insight towards the development of sound event recognition which able to cope with label noise and minimal supervision conditions. Instead of manually annotation, labels can be inferred automatically mainly by two kinds of methods: applying pre-trained model on the audio contents and using heuristic of the metadata. For example, variants of VGG \cite{simonyan2014vgg} and ResNet \cite{ResNet} are trained on a large collection of YouTube videos, then automatically annotating the audio tracks from Flickr videos \cite{Fonseca2019audio}. The noisy labels are introduced due to the bias of the pre-trained model. 
The warm-up pipeline \cite{Bouteillon2019} contains 4 stages, each stage generates one model which is trained on a different curated or noisy subset. The parameters of model in the previous stage are used to initialize the model in the next stage. It can be regarded as a kind of ensemble learning to deal with the noisy label problem for audio tagging.

Webly-labeled data are common type of noisy labeled data, they could be collected through searching from the websites (e.g. text query on YouTube). The labels are mainly based on the metadata maintained by users or websites. This kind of webly-labeled data may introduce noisy labels because of two causes: 1) The impreciseness of event names or low-quality metadata of videos on the websites. 2) Search engines like Google usually operate in the high-precision low-recall regime.
WeblyNet \cite{kumar2018learning} addresses this noisy data issue by using webly-labeled data. It also maintains two networks and applies the multi-view learning. The two networks are trained jointly by adding a penalty term to the losses of two views.

\section{Method}
\label{sec:method}

In this section, we present CrossFilter, a robust learning framework under noisy labeled data for audio tagging. For the bi-quality data, we denote a small size curated data set as $\mathcal{C}=\{(x_c, y_c)|c=1,2,...,C\}$ with $C$ manually verified instances and a large-scale noisy labeled data set $\mathcal{N}=\{(x_n, y_n)|n=1,2,...,N \}$ with $N$ corrupted instances. The samples in $\mathcal{C}$ is highly credible and samples in $\mathcal{N}$ may have incorrect labels. Our goal is to learn a model from training set $\mathcal{C} \cup \mathcal{N}$ to classify unseen instances.

The overview of our approach is shown in Figure \ref{fig:overview}. It consists of three main components: 
1) Cross-representation. We take different audio representations to train a couple of networks $M^1$ and $M^2$. Complementary audio representations increase the divergence between the two classifiers. To our knowledge, this is the first work to use multiple audio representations to cope with label noise.
2) Noise filtering. We employ two networks to select possibly correct data for training. Noise Filtering algorithm is proposed to select samples from the noisy subset $\mathcal{N}$ for the peer network. 
It leads the networks to learn to distinguish whether a sample is correctly or incorrectly labeled. If the sample is correctly labeled,  then it is used to extend the $\mathcal{C}$ subset.
3) Multi-task learning. As we cannot accurately select out all the clean data from the noisy subset, the remaining samples in $\mathcal{N}$ may contain valuable samples. To tackle this problem, we investigate to improve the generalization performance though multi-task learning (MTL) which employs disparate criterions and classifiers for $\mathcal{C}$ and $\mathcal{N}$.

\subsection{Cross-Representation}

The performance of audio analysis is highly depended on the representations of the audio signals. The widely used representations include: Spec, Logmel, MFCC and CQT etc. These representations are presented in two dimensional $time$-$frequency$ (or $time$-$quefrency$) format.

Different representations describe the divergence characteristics of audio. For example, in the linear-frequency power spectrogram (Spec), the scale of power value is linear. Relatively, Spec emphasizes the information of medium and high frequency range, while the other representations (MFCC, CQT and Logmel) pay more attention to the low frequency (or quefrency) range. Logmel is obtained by applying logarithm and mel-scale on the spectrograms which been transformed from wave by Discrete Fourier Transformation (DFT) \cite{virtanen2018computational}. MFCC is obtained by applying the Discrete Cosine Transformation (DCT) on a log-mel spectrogram and it could provide a more compact and more smooth representation compared with Logmel \cite{young1993htk}. Different from DFT, the ratio of center frequency to resolution is a constant value in CQT \cite{brown1991calculation,seneff1985pitch}. It yields better results where low frequencies are concerned. 
Previous works \cite{wilkinghoff2018general,weiping2017} used multiple audio representations to improve the performance by feature fusion or prediction ensemble. In our approach, cross-representations is used for robust learning under noisy labeled data. Models trained with different representations could learn complementary information of audio signals and introduce the dissimilarity among these models.

For the bi-quality dataset, $\{\mathcal{C}^1, \mathcal{N}^1\}$ and $\{\mathcal{C}^2, \mathcal{N}^2\}$ are two disjoint partitions of whole dataset. Before training, the two partitions are equal. In the follow-up training process, Logmel transformation ($Rep^1$) would be conducted on $\{\mathcal{C}^1, \mathcal{N}^1\}$ and CQT transformation ($Rep^2$) would be conducted on $\{\mathcal{C}^2, \mathcal{N}^2\}$. In experiment section, we invest the effects of cross-representation for classification performance.

\begin{algorithm}[t]
\caption{Noise Filtering}
\label{alg:NF}
\begin{algorithmic}[1]

\REQUIRE ~~\\
Original partition of dataset $\{\mathcal{C}, \mathcal{N}\}$. \\
$Rep^1(x)$ and $Rep^2(x)$ give two representations of audio instance $x$. \\
Agree($a, b$): The most confident class in predicted probability $b$ is consistent with target label $a$.\\
$\delta(y_n, \mathcal{C}^r)$ indicates the number of pseudo curated data with label $y_n$ in $\mathcal{C}^r$
\STATE{\textbf{Initialization:}}

\STATE{$\mathcal{C}^1 = \mathcal{C}^2 = \mathcal{C}, \mathcal{N}^1 = \mathcal{N}^2= \mathcal{N}, k = 0$}

\STATE{\textbf{Training:}}
\FOR{$j=1$ to $EPOCH$}
  \STATE {Train $M^1$ and $M^2$ with mini-batch from $(\mathcal{C}^1, \mathcal{N}^1)$ and $(\mathcal{C}^2, \mathcal{N}^2)$ respectively.}
    \STATE {$\mathcal{N} =$ Shuffle($\mathcal{N}$)}
    \FORALL{$(x_n, y_n) \in \mathcal{N}$}

      \STATE{$\hat{y_n^1} = M^1(Rep^1(x_n))$}
      \IF{Agree($y_n, \hat{y_n^1}$) and $\delta(y_n, \mathcal{C}^1) < k$}
        \STATE{$\mathcal{N}^2 = \mathcal{N} - \{(x_n, y_n)\}$}
        \STATE{$\mathcal{C}^2 = \mathcal{C} \cup \{(x_n, y_n)\}$}
      \ENDIF

      \STATE{$\hat{y_n^2} = M^2(Rep^2(x_n))$}
      \IF{Agree($y_n, \hat{y_n^2}$) and $\delta(y_n, \mathcal{C}^2) < k$} 
        \STATE{$\mathcal{N}^1 = \mathcal{N} - \{(x_n, y_n)\}$}
        \STATE{$\mathcal{C}^1 = \mathcal{C} \cup \{(x_n, y_n)\}$}
      \ENDIF
    \ENDFOR
  \STATE{$k = $Step$(j)$}
\ENDFOR
\RETURN{$M^1, M^2$}
\end{algorithmic}
\end{algorithm}

\subsection{Noise Filtering}

By learning with cross-representation, we hope to select the noisy samples and re-partition $\{\mathcal{C}^r, \mathcal{N}^r\}, (r=1,2)$ by the cooperation of two peer networks. The $\mathcal{C}^r$ is expanded by selecting clean data from $\mathcal{N}$. Here we propose the Noise Filtering algorithm (NF) to complete this selecting process. Pseudo code is given in Algorithm \ref{alg:NF}.

Specifically, we train network $M^1$ using Logmel representation and train $M^2$ with CQT representation. For $(x_n, y_n)$ in $\mathcal{N}^1$, if the prediction for $x_n$ with network $M^1$ is consistent with the noisy label, we might consider that this noisy label $y_n$ is correct and we named it pseudo curated data. Then we move the instance $(x_n, y_n)$ from $\mathcal{N}^2$ to $\mathcal{C}^2$ at the next epoch. The movement is conducted on $\{\mathcal{C}^2, \mathcal{N}^2\}$ instead of $\{\mathcal{C}^1, \mathcal{N}^1\}$, because we hope to avoid accumulated error caused by sample-selection bias. In this way, the error from one network will not be directly transferred back to itself, but been revised by peer network which is trained under a different representation. Similarly, if the prediction of the sample through the network $M^2$ is the same as the given label, we remove this sample from $\mathcal{N}^1$ and add it to $\mathcal{C}^1$. Thus, we use a network to filter the noisy data for another network, and in this way the performance of different networks is promoted via cross representation.

Inspired by the curriculum learning \cite{bengio2009curriculum,jiang2015self}, we gradually increase the number of pseudo curated data selected from $\mathcal{N}$ in each epoch. 
The $\delta(y_n, \mathcal{C}^r)$ indicates the number of pseudo curated data with label $y_n$ in $\mathcal{C}^r$. 
In order to avoid making $\mathcal{C}^r$ severely imbalance on category, we randomly select at most $k$ samples for every predicted class by $\delta(y_n, \mathcal{C}^r) < k$. 
The upper bound of the newly added pseudo curated data is $min(N, J\times k)$, where $J$ is the number of classes. With the training process, we gradually increase the value of $k$ with Step$(j)$. For example, we increase the $k$ linearly as the epoch grows. The network can improve the generalization ability by using larger $\mathcal{C}^r$.

\subsection{Multi-Task Learning}

To minimize the disturbance of wrong labels, the previous approaches discard $\mathcal{N}^r$ and use $\mathcal{C}^r$ to train the network \cite{malach2017decoupling,yu2019does}. However, we can not accurately select out all the clean data in $\mathcal{N}$. If we only use $\mathcal{C}^r$ for training, the dataset will be much smaller which is prone to over-fitting. In order to take advantage of bi-quality datasets and reduce the adverse effects of incorrect labels, we propose to use multi-task learning on $\mathcal{C}^r$ and $\mathcal{N}^r (r=1,2)$. 

Under the MTL framework, we set different classifiers $\bm{f}(\cdot; \bm{\theta_c})$ and $\bm{f}(\cdot; \bm{\theta_n})$ for $\mathcal{C}^r$ and $\mathcal{N}^r$ respectively. $\bm{\theta_c}$ and $\bm{\theta_n}$ are the parameters of the two classifiers. They classify the features extracted by the same backbone network. The criterion for $\mathcal{C}^r$ and $\mathcal{N}^r$ is denoted as $loss_c$ and $loss_n$ respectively. The sample specific loss depends on the current instance is from whether $\mathcal{C}^r$ or $\mathcal{N}^r$. For curated set $\mathcal{C}^r$, $loss_c$ is categorical cross-entropy loss (CCE) for single-label classification and binary cross-entropy (BCE) for multi-label classification:

\begin{equation}
\label{eq:CCE}
loss_{c(CCE)} = -\sum_{j=1}^{J}\bm{y}_{ij}\log(\phi_j(\bm{f}(\bm{x}_i;\bm{\theta_c})),
\end{equation}
\begin{equation}
\label{eq:BCE}
\begin{split}
loss_{c(BCE)} = & -\sum_{j=1}^{J}\bm{y}_{ij}\log(\sigma_j(\bm{f}(\bm{x}_i;\bm{\theta_c}))) \\& -\sum_{j=1}^J(\bm{1}-\bm{y}_{ij})\log(\sigma_j(\bm{1}-\bm{f}(\bm{x}_i;\bm{\theta_c}))),
\end{split}
\end{equation}

where $J$ is the number of classes, $i$ is the index of audio samples. $\bm{y}_{ij}$ is the $j$'th element of one-hot encoded label. $\phi_j$ and $\sigma_j$  denote the $j$'th element of Softmax function and Sigmoid function respectively.

For the auxiliary task which carries out with noisy labels, we employ the noisy-robust loss function \cite{fonseca2019learning} $L_q$ as $loss_n$:
\begin{equation}
loss_n = \frac{1-(\sum_{j=1}^J\bm{y}_{ij}\phi_j(\bm{f}(\bm{x}_i;\bm{\theta_n})))^q}{q},
\end{equation}

where $\quad q\in(0,1]$. It can be proved that $L_q$ becomes mean absolute error (MAE) when $q=1$ and $\lim_{q\to0}{L_q}$ is equivalent to cross-entropy loss (CCE) \cite{zhang2018generalized}. Therefore, the $L_q$ is a generalization of CCE and MAE. 
For more details, CCE suffers from the wrong labels, because CCE is weighted more for the gradient update if the predictions differ more from the target label. This is undesirable in the case of noisy label. On the other hand, it is theoretically proved \cite{ghosh2017robust} that MAE is robust against noisy label because MAE weights all the prediction equally. But the derivative of MAE is not continuous and it is hard to optimize under high-dimensional situation. In the image and audio classification task, the MAE often takes significantly longer to converge and brings performance degradation. Benefiting from both CCE and MAE, the $loss_n$ is noise-robust and easy to be optimized for audio data.  The risk on the bi-quality dataset $\mathcal{C} \cup {\mathcal{N}}$ becomes:
\begin{equation}
\mathcal{R}_{\mathcal{C} \cup {\mathcal{N}}} = \mathbb{E}_{\mathcal{C}}[loss_c] + \lambda\mathbb{E}_{\mathcal{N}}[loss_n]
\end{equation}

where $\mathbb{E}$ denotes the expectation over the training samples. The hyper-parameter $\lambda$ can be set through a grid search on validation set. Note that, in the Noise Filtering phase, we assign single pseudo-label to every noisy instance. Thus, we employ the same $loss_n$ for both single-label and multi-label classification tasks. In experiment section, we compare different criterion setting for $loss_c$ and $loss_n$.

\subsection{Inference}
When evaluating our model on the unseen test data, we randomly cut out 5 audio segments of 4-seconds length from each audio instance. If the length is not enough, zeros are padded at start and end positions. The results of each instance are predicted by averaging the probability of 5 segments.
The classifier trained on $\mathcal{C}^r$ is more reliable than classifier trained on $\mathcal{N}^r$ because the supervision signal is more credible. So we only use the curated data path when predicting new data. We simply add the output probability for the two models trained with cross-representation. There might be more effective ways to ensemble the output of the two networks, but that is not the focal point of our research.

\subsection{Relations to Previous Works}
\label{sub:relations_to_previous_works}
We compare our work with two other works that also use the idea of ``co-learning'': Co-teaching \cite{han2018co} and WeblyNet \cite{kumar2018learning}.

Co-teaching trains two neural networks simultaneously. It learns from the noisy labeled data in following steps: each network feeds forward all data and selects the data with small-loss (possibly clean labels), then each network back propagates the mini-batch data selected by its peer network and updates itself. In this method, it is important to keep the diversity of the two networks. However, the diversity of the two networks in Co-teaching depends on the different weights initialization. They gradually converge to consensus in the training process, and the mini-batch losses of two networks turn to close. It leads to the inability for selecting correctly labeled data. 
In our proposed work, training two networks with different representations further increases the diversity of peer networks. Further, Co-teaching algorithm needs the ratio of noisy labels (incorrect labels as a percentage of all labels) which is usually unknown for many practical dataset. While, the ratio of noisy labels is not necessary for the CrossFilter.

WeblyNet also maintains two networks and applies the multi-view learning. One network is a deep CNN and another is a 3 layers full-connected network. The second network uses the bottleneck feature of pre-trained model as input. The two networks are trained jointly by adding a penalty term (generalized KL-divergence \cite{banerjee2005clustering}) to the BCE losses of two views. However, the ``representation'' in our approach is essentially different from the ``view'' in WeblyNet. Different views are two kinds features extracted by different networks with the same representation. This intrinsically determines that the two networks are learning from the same materials. However, the representations vary greatly according to different transformations. 
Besides, both Co-teaching and WeblyNet may suffer from the class-imbalanced issue which is common in the webly-labeled data. With the proposed NF algorithm, CrossFilter is trained in the class-balanced way. 
Moreover, in order to obtain distinctive features, one of the peer networks in WeblyNet uses large amount of external data (AudioSet \cite{crocco2016audio}) for pre-training. It is reasonable to get a competitive performance at the cost of more training data and time. Our framework is trained from scratch only on the experimental data. This may place CrossFilter at an unfavorable situation. Even so, CrossFilter is competitive.

Section \ref{sec:Experiments} empirically compares these three works.

\section{Experiments Procedures}
\label{sec:Experiments}
\subsection{Datasets}

We employ two widely-used bi-quality audio tagging datasets, FSDKaggle2018 \cite{Fonseca2018_DCASE}, and FSDKaggle2019 \cite{Fonseca2019audio} to train and evaluate our models. 

\textbf{FSDKaggle2018}. The train set includes about 9.5k clips with 41 categories. The audio clips are obtained from Freesound content annotated with labels from AudioSet ontology \cite{gemmeke2017audio}. The duration of the audio samples ranges from 300ms to 30s due to the diversity of the sound instances. A single label is assigned to each audio clip. The dataset is bi-quality, which means the train set is composed of about 3.7k curated annotations and about 5.8k noisy annotations. The quality of the noisy annotations has been roughly estimated to be at least 65-70\% in each sound category. The test set is composed of 1.6k manually-verified annotations with a similar category distribution of the train set. 

\textbf{FSDKaggle2019}. This dataset is released on the DCASE 2019 challenge. Different from FSDKaggle2018, this dataset is under a large vocabulary (80 categories) and about 20\% instances in this dataset have multiple tags. The labels also come from the AudioSet ontology. It is also a bi-quality dataset which consists of 10.5 hours (about 5k instances) curated data and about 80 hours (about 19.8k instances) noisy labeled data. The audio clips in the curated subset are from Freesound and the audio clips in the noisy subset are from the sound tracks of a pool of Flickr videos. This introduces a potential domain mismatch. The test set is composed of about 4.5k manually-verified data from the same source of the curated subset. 

\begin{table}[t]
\setlength{\tabcolsep}{8pt}
\setlength\extrarowheight{1.5pt}
\centering
\caption{The overall results of CrossFilter and other competing approaches on two datasets: FSDKaggle2018 and FSDKaggle2019.}
\label{tb:overall}
\begin{tabular}{lcc}
\toprule[1pt]
METHOD                 & \begin{tabular}[c]{@{}c@{}}FSD-2018\\ mAP@3(\%)\end{tabular} & \begin{tabular}[c]{@{}c@{}}FSD-2019 \\ lwlrap\end{tabular} \\
\midrule
Baseline\cite{Fonseca2018_DCASE,Fonseca2019audio}   & 69.43     & 0.5460   \\
MTL+Self supervised\cite{lee2019label}           & 72.60     & -      \\
Cross-task Learning\cite{kong2018dcase} & 90.30     & -        \\
Pseudo-Label\cite{lee2013pseudo}           & 91.52     & 0.6883        \\
Surrogate Loss\cite{fonseca2019learning}       & 90.87     & 0.6531      \\
WeblyNet\cite{kumar2018learning}   & 84.67 & 0.6172 \\
Co-teaching\cite{han2018co}           & 92.50     & 0.7071   \\
Iterative Training\cite{Nguyen2018}\scriptsize{(ensemble)}           & 94.96     & -      \\
Loss Masking\cite{jeong2018audio}\scriptsize{(ensemble)}            & 95.38     & -      \\
\midrule
\textbf{Ours}                   & \textbf{95.59}     & \textbf{0.7195}    \\
\bottomrule[1pt]
\end{tabular}
\end{table}

\subsection{Evaluation Metrics}
\label{sub:metrics}
1) For the FSDKaggle2018 dataset, accuracy and mean average precision at cutoff 3 (mAP@3) are evaluated to keep consistent with previous works \cite{Fonseca2018_DCASE,kong2018dcase,iqbal2018general}. Formally, the accuracy is defined as:

\begin{equation}
Accuracy = \frac{TP+TN}{U},
\end{equation}

where $U$ is the number of scored audio files in the test data, true positive (TP), true negative (TN) are basic statistics recording the correspondence between given labels and predictions. TP refers to both the prediction and ground truth label indicate the presence of a sound event in the audio recording, TN refers to both the prediction and ground truth label indicate the absence of a sound event. The $mAP@3$ is defined as:
\begin{equation}
mAP@3 = \frac{1}{U} \sum_{u=1}^{U} \sum_{k=1}^{min(n,3)} P(k),
\end{equation}
where $n$ is the number of predictions per audio clip and $P(k)$ is the precision at cutoff $k$.

2) For the FSDKaggle2019 dataset, some audio clips bear one label while others bear several labels.
The task consists of predicting the audio labels (tags) for every test clip. Some test clips bear one label while others bear several labels. To evaluate the multi-label audio tagging, we use the label-weighted label-ranking average precision (lwlrap) as the primary metric which is also suggested by the DCASE 2019 audio tagging challenge \cite{Fonseca2019audio,Bouteillon2019}.

Formally, $n_{clips}$ and $n_{classes}$ are the numbers of audio clips and classes respectively. Given a binary indicator matrix of the ground truth labels $y \in \{0, 1\}^{n_{clips} \times n_{classes}}$ and the predicted score matrix $\hat{f} \in \mathbb{R}^{n_{clips} \times n_{classes}}$, the lwlrap score can be computed as:

\begin{equation}
lwlrap = \frac{1}{||y||_0}\sum_{c=1}^{n_{classes}}\sum_{i:y_{ic}=1}\frac{1}{||y_{i}||_0}\sum_{j:y_{ij}=1}\frac{|\mathcal{L}_{ij}|}{rank_{ij}},
\end{equation}
where
$$\mathcal{L}_{ij} = \{k:y_{ij}=1, \hat{f}_{ik} \geqslant \hat{f}_{ij}\},$$
$$rank_{ij} = |\{k:\hat{f}_{ik} \geqslant \hat{f}_{ij}\}|,$$
the $|\cdot|$ computes the cardinality of the set (the number of element in the set), 
and $||\cdot||_0$ is the $\ell_0$ "norm", which computes the number of nonzero elements in a matrix or vector.
This measures the average precision of retrieving a ranked list of relevant labels for each test clip. It will be higher if one is able to give better rank to the labels associated with each sample. The obtained score is always strictly greater than 0, and the best value is 1. The "label-weighted" part means that the overall score is the average over all the labels in the test set, where each label receives equal weight.

\begin{table*}[!t]
\setlength{\tabcolsep}{3.4pt}
\setlength\extrarowheight{2pt}
\centering
\caption{The mAP@3 scores among the Co-teaching, WeblyNet and proposed CrossFilter of all the classes on FSDKaggle2018.}
\label{tb:class_performance}
\begin{tabular}{ccccccccccccccc}
\toprule[1pt]
& \begin{tabular}[c]{@{}c@{}}Acous.\\ guitar\end{tabular} & \begin{tabular}[c]{@{}c@{}}Appl-\\ ause\end{tabular} & Bark & \begin{tabular}[c]{@{}c@{}}Bass\_\\ drum\end{tabular} & Burping  & Bus   & Cello  & Chime & Clarinet & \begin{tabular}[c]{@{}c@{}}Computer\_\\ keyboard\end{tabular} & Cowbell & Drawer & Fart & \begin{tabular}[c]{@{}c@{}}Finger\_\\ snap\end{tabular} \\
\hline
Co-teaching\cite{han2018co} & 87.04            & 100.0   & 96.43        & 96.43      & 100.0                  & 76.67          & 92.59        & 82.18           & 98.21                & 91.67              & 98.81    & 78.16                   & 93.89              & 97.98            \\
WeblyNet\cite{kumar2018learning}    & 48.15            & 79.17    & 100.0       & 100.0     & 96.35                   & 96.00          & 96.30        & 82.76           & 95.24                & 91.67              & 92.86    & 76.44                   & 74.44              & 95.45            \\
\textbf{CrossFilter} & 84.81            & 100.0   & 98.21        & 96.43      & 100.0                  & 94.00          & 97.30        & 89.66           & 100.0               & 96.15              & 97.62    & 90.80                   & 100.0             & 98.48            \\
\hline\hline
& \begin{tabular}[c]{@{}c@{}}Fire-\\ works\end{tabular} & Flute & \begin{tabular}[c]{@{}c@{}}Glock-\\ enspiel\end{tabular} & Gong & \begin{tabular}[c]{@{}c@{}}Harm-\\ onica\end{tabular} & Keys & Knock & \begin{tabular}[c]{@{}c@{}}Lau-\\ ghter\end{tabular} & Meow & \begin{tabular}[c]{@{}c@{}}Microwave\_\\ oven\end{tabular} & Oboe & \begin{tabular}[c]{@{}c@{}}Saxo-\\ phone\end{tabular}  & Shatter & Squeak\\
\hline
Co-teaching\cite{han2018co} & 71.35            & 94.55    & 89.66        & 97.30      & 95.45                   & 83.93          & 93.16        & 97.37           & 93.10                & 89.66              & 100.0   & 98.18                   & 100.0             & 49.43            \\
WeblyNet\cite{kumar2018learning}    & 59.90            & 86.97    & 64.94        & 82.88      & 37.37                   & 87.50          & 91.03        & 100.0          & 98.28                & 63.79              & 97.22    & 95.45                   & 86.78              & 37.36            \\
\textbf{CrossFilter} & 69.79            & 98.79    & 83.03        & 97.30      & 98.48                   & 92.86          & 96.58        & 96.05           & 98.28                & 96.55              & 98.81    & 99.39                   & 100.0             & 71.26            \\
\hline\hline
& \begin{tabular}[c]{@{}c@{}}Tamb-\\ ourine\end{tabular}  & Tearing & \begin{tabular}[c]{@{}c@{}}Tele-\\ phone\end{tabular} & \begin{tabular}[c]{@{}c@{}}Tru-\\ mpet\end{tabular}  & Writing & Cough & \begin{tabular}[c]{@{}c@{}}Double\_\\ bass\end{tabular} & \begin{tabular}[c]{@{}c@{}}Electric\_\\ piano\end{tabular} & \begin{tabular}[c]{@{}c@{}}Gunshot\_\\ gunfire\end{tabular} & Hi-hat & Scissors & \begin{tabular}[c]{@{}c@{}}Snare\_\\ drum\end{tabular} & \begin{tabular}[c]{@{}c@{}}Violin\_\\ fiddle\end{tabular}  & \textbf{Avg.}\\
\hline
Co-teaching\cite{han2018co} & 97.50            & 87.65    & 84.03        & 94.14      & 87.36                   & 96.67          & 100.0       & 94.79           & 91.01                & 97.44              & 62.67    & 100.0                  & 99.38              &     92.95              \\
WeblyNet\cite{kumar2018learning}    & 89.17            & 72.22    & 84.72        & 76.58      & 71.26                   & 93.33          & 90.83        & 93.23           & 93.39                & 70.09              & 88.67    & 100.0                  & 97.07              &     85.49             \\
\textbf{CrossFilter} & 95.00            & 98.15    & 89.54        & 95.95      & 90.80                   & 98.33          & 100.0       & 100.0          & 95.77                & 94.44              & 93.33    & 100.0                  & 100.0             &      95.59      \\      
\bottomrule[1pt]
\end{tabular}
\end{table*}

\begin{table*}[!t]
\setlength{\tabcolsep}{2.2pt}
\setlength\extrarowheight{4pt}
\centering
\caption{The best and worst 15 classes on the test set of FSDKaggle2019.}
\label{tb:best_and_worst}
\begin{tabular}{cccccccccccccccc}
\toprule[1pt]
& Sneeze & \begin{tabular}[c]{@{}c@{}}Computer-\\ keyboard\end{tabular} & Purr   & Zipper & \begin{tabular}[c]{@{}c@{}}Burping-\\ eructation\end{tabular} & Shatter  & \begin{tabular}[c]{@{}c@{}}Keys-\\ jangling\end{tabular} & \begin{tabular}[c]{@{}c@{}}Bicycle-\\ bell\end{tabular} & \begin{tabular}[c]{@{}c@{}}Bass-\\ drum\end{tabular} & Drawer                                                   & Meow                                                       & \begin{tabular}[c]{@{}c@{}}Microwave-\\ oven\end{tabular}   & Applause                                               & Writing                                                & \begin{tabular}[c]{@{}c@{}}Church-\\ bell\end{tabular} \\
\hline
lrap   & 0.9741 & 0.9660                                                       & 0.9543 & 0.9528 & 0.9422                                                        & 0.9380   & 0.9375                                                   & 0.9344                                                  & 0.9335                                               & 0.9211                                                   & 0.9179                                                     & 0.9107                                                      & 0.9101                                                 & 0.8999                                                 & 0.8831                                                 \\
weight & 0.0093 & 0.0102                                                       & 0.0099 & 0.0173 & 0.0152                                                        & 0.0101   & 0.0118                                                   & 0.0093                                                  & 0.0138                                               & 0.0115                                                   & 0.0118                                                     & 0.0112                                                      & 0.0171                                                 & 0.0155                                                 & 0.0093     \\
\hline\hline
& Bus    & \begin{tabular}[c]{@{}c@{}}Male-\\ singing\end{tabular}      & Car    & Buzz   & \begin{tabular}[c]{@{}c@{}}Child-\\ speech\end{tabular}       & Cupboard & Drip                                                     & Squeak                                                  & Gurgling                                             & \begin{tabular}[c]{@{}c@{}}Female-\\ speech\end{tabular} & \begin{tabular}[c]{@{}c@{}}Trickle-\\ dribble\end{tabular} & \begin{tabular}[c]{@{}c@{}}Dishes-\\ pots\end{tabular} & \begin{tabular}[c]{@{}c@{}}Male-\\ speech\end{tabular} & \begin{tabular}[c]{@{}c@{}}Chirp-\\ tweet\end{tabular} & Tap                                                    \\
\hline
lrap   & 0.5513 & 0.5318                                                       & 0.5229 & 0.4983 & 0.4972                                                        & 0.4824   & 0.4820                                                   & 0.4778                                                  & 0.4563                                               & 0.4327                                                   & 0.3638                                                     & 0.3629                                                      & 0.3063                                                 & 0.3052                                                 & 0.2217                                                 \\
weight & 0.0123 & 0.0082                                                       & 0.0162 & 0.0080 & 0.0114                                                        & 0.0091   & 0.0184                                                   & 0.0117                                                  & 0.0205                                               & 0.0181                                                   & 0.0240                                                     & 0.0155                                                      & 0.0240                                                 & 0.0118                                                 & 0.0227          \\                                      
\bottomrule[1pt]
\end{tabular}
\end{table*}

\subsection{Implementation Details}
We use MobileNetV2\cite{sandler2018mobilenetv2} as our backbone network which is same as the DCASE baseline system. The width multiplier in MobileNetV2 is set to 1. This lightweight backbone is computational efficient and other heavy backbone might lead to a better result at a cost of computational resources. The global mean-max pooling is followed after the backbone networks. The representation dimension on $time$ axis is usually larger than that on $frequency$ (or $quefrency$), for example, the dimensions of our input is $64\times800$. To emphasize the frequency range with the highest value, we use global max pooling on $frequency$ (or $quefrency$). Meanwhile, considering the amplitude of many sound events fluctuate in the time dimension, we employ the global mean pooling on $time$.


Pre-processing of audio in our experiments is conducted with uncompressed PCM 16 bits and 44.1 kHz mono audio format. Librosa is used for the audio pre-processing. In the Spec, Logmel and MFCC representations, we use the same frame width with 100ms and frame shift with 5ms. The hop length for CQT is 256 (5.8ms frame shift), which is close to other three kinds of representations. The number of frequency bins for Logmel, MFCC and CQT is set to 64. No truncation is used for MFCC bins. For the Power spectrogram, the dimension of frequency axis is much larger than other representations (It is determined by the frame width). To keep the same size of representations in the experiments, we use the mean pooling to reduce the dimension of the frequency axis to 64. We randomly crop a 4-seconds segment from an audio clip at every epoch during training. SpecAugment \cite{park2019specaugment} and MixUp \cite{xu2018mixup} are applied for data augmentation. In SpecAugment, one frequency masking and one time masking is used and they are less than 10\% and 20\% of the maximum width respectively. For the trick of mixup, mixing ratio of sample pairs $\gamma \sim Beta(\alpha, \alpha)$, where $\alpha = 1$ for all the experiments. In the testing phase, no data augmentation is used. 

Adam optimizer is used to optimize our loss and all weight parameters are subjected to $\ell_2$ regularization with coefficient $5 \times 10 ^ {-6}$. The Cosine Annealing Learning Rate with warmup is used as the learning rate scheduler. More specifically, the learning rate linearly grows from $5 \times 10^{-5}$ to $5 \times 10^{-4}$, then gradually anneal to $5 \times 10^{-6}$ in 300 epochs. 
To make the results more convincing, we use the stratified 5-folds to validate our model and report the mean performance on the 5-folds.

\subsection{Compared Methods}
As comparison, the results of Baseline \cite{Fonseca2018_DCASE,Fonseca2019audio}, Cross-task \cite{kong2018dcase}, Iterative Training \cite{Nguyen2018} and Loss masking \cite{jeong2018audio} are reported in the literature. Meanwhile, we have reproduced some of the most common and up-to-date methods in the field. 
In the Pseudo-label, we firstly pre-train the peer network by $\mathcal{C}$ and make predictions for the instances in $\mathcal{N}$. Then the network is fine-tuned on $\mathcal{C}$ with clean labels and $\mathcal{N}$ with predicted labels. 
In the Surrogate Loss approach \cite{fonseca2019learning}, $L_q$ is applied as the noisy-robust objective criterion. The $q$ is set as 0.7 which give the best performance on the entire FSDnoisy18k dataset. The data source of FSDnoisy18k is same as our experimental dataset and the classes are all included in our experiments. It is reasonable to use the same hyper-parameter. 
For Co-teaching approach, as noisy ratio $\epsilon$ is unknown and it is set to 0.3 for the dataset according to the official estimation of noisy ratio \cite{Fonseca2018_DCASE}. 
The WeblyNet approach is proposed to solve the Sound Event Detection (SED) task \cite{kong2019sound,adavanne2018sound} which predicts the onset and offset time of sound events. 
In order to better adapt to the FSDKaggle2018 dataset, which is a single-label tagging task, the sigmoid activation function is replaced by Softmax operation and BCE loss is replaced by CCE loss. For all the reproduction, we use Logmel representation and the data augmentations are kept the same as our proposed method.
Most of the hyper-parameters in the compared approaches are configured the same as our method, including data augmentation hyper-parameters, optimizer type and weight decay. Further, we re-tune the hyper-parameters for Co-teaching and WeblyNet on our experimental data. In the Co-teaching, the initial learning rate is $1\times10^{-4}$ then anneal to $1\times10^{-5}$, and the optimizer momentum is 0.9 as original paper used. For WeblyNet, in order to update the two different models effectively, we use different learning rates for two networks, $1\times10^{-3}$ for the full-connect network and $1\times10^{-4}$ for the CNN. Also, they gradually decay 10 times during training.

\begin{table}[t]
\centering
\setlength{\tabcolsep}{8pt}
\setlength\extrarowheight{2pt}
\caption{The performances under different representations. For the single representation, Logmel gives the best results. For the cross-representation, the combination of Logmel and CQT performs best. Evaluation on FSDKaggle2018 dataset.}
\label{tb:exp_cr}
\begin{tabular}{c|c|cc}
\toprule[1pt]
                            & Representation & Accuracy   & mAP@3      \\
\midrule
\multirow{4}{*}{\begin{tabular}[c]{@{}c@{}}Single\\Rep\end{tabular}} & Spec+Spec           & 84.26\tiny{$\pm$0.25} & 88.63\tiny{$\pm$0.28}  \\
                            & MFCC+MFCC          & 85.44\tiny{$\pm$0.31} & 89.62\tiny{$\pm$0.21} \\
                            & Logmel+Logmel         & \textbf{87.85}\tiny{$\pm$0.79} & \textbf{91.53}\tiny{$\pm$0.46} \\
                            & CQT+CQT            & 87.35\tiny{$\pm$0.54} & 90.87\tiny{$\pm$0.40} \\
\midrule
\multirow{6}{*}{\begin{tabular}[c]{@{}c@{}}Cross\\Rep\end{tabular}}  & Spec+MFCC      & 86.11\tiny{$\pm$0.28} & 90.79\tiny{$\pm$0.21} \\
                            & Spec+Logmel    & 87.82\tiny{$\pm$0.38} & 91.38\tiny{$\pm$0.30} \\
                            & Spec+CQT       & 87.41\tiny{$\pm$0.47} & 91.25\tiny{$\pm$0.31} \\
                            & MFCC+Logmel    & 88.26\tiny{$\pm$0.76} & 91.69\tiny{$\pm$0.43} \\
                            & MFCC+CQT       & 88.34\tiny{$\pm$0.59} & 91.84\tiny{$\pm$0.37} \\
                            & Logmel+CQT     & \textbf{89.11}\tiny{$\pm$0.54} & \textbf{92.43}\tiny{$\pm$0.33} \\
\bottomrule[1pt]
\end{tabular}
\end{table}

\subsection{Overall Results}
\label{sub:exp_overall}

Table \ref{tb:overall} shows the overall results of CrossFilter and other competing approaches on two experimental datasets. Our approach achieves 95.59\% mAP@3 on FSDkaggle2018, which is currently the state-of-the-art result even compared with other ensemble approaches on the leaderboard. On the latest FSDKaggle2019 dataset, we achieve 0.7195 lwlrap score, which have a great improvement compared with other related works.

Table \ref{tb:class_performance} shows the mAP@3 of all the sound events on the FSDKaggle2018 dataset. We compare with the most relevant works, Co-teaching and WeblyNet. The classes such as \textit{Chime}, \textit{Computer\_keyboard}, \textit{Keys}, \textit{Squeak} and \textit{Scissors} have obvious improvements compared with other two approaches. Most of these sound events are transient and short-lasting audio instances. The network is hard to distinguish them effectively using single representation. The samples of these kinds of classes are more likely to be incorrectly labeled because they are also hard to be recognized by humans. Also, the total length of these classes is shorter than other durative classes like music and speech. So, the influence of noisy label on these categories is more serious. On the other hand, the improvements on the classes such as \textit{Laughter}, \textit{Meow}, \textit{Oboe}, \textit{Saxophone} and \textit{Telephone} are more moderate. These classes are usually composed of continuous and durative audio instances. Table \ref{tb:best_and_worst} gives the 15 best and worst performing classes on FSDKaggle2019 and their label weights. The results are given on the entire test set. The results suggest our approach could accurately tag on classes like \textit{Sneeze}, \textit{Computer-keyboard} and \textit{Purr}. However, some classes like \textit{Tap} and \textit{Chirp} are difficult to recognize under a large vocabulary setting (80 classes). More insight into error analysis is given in the Section \ref{sec:conclusion}.

\begin{table}[t]
\centering
\setlength{\tabcolsep}{8pt}
\caption{mAP@3 score on FSDKaggle2018 with or without noise filtering (NF).}
\label{tb:nf_on_2018}
\begin{tabular}{cccc}
\toprule[1pt]
           & $M^1$(Lomgel) & $M^2$(CQT) & $M^1 + M^2$ \\
\midrule
Without NF & 93.04\tiny{$\pm$0.40} & 92.87\tiny{$\pm$0.54}  & 94.08\tiny{$\pm$0.22}   \\
NF         & \textbf{94.68}\tiny{$\pm$0.17} & \textbf{94.50}\tiny{$\pm$0.38}  & \textbf{95.59}\tiny{$\pm$0.20}    \\
\bottomrule[1pt]
\end{tabular}
\end{table}
\section{Results}
\label{sec:Results}

\begin{table}[t]
\centering
\setlength{\tabcolsep}{8pt}
\caption{lwlrap score on FSDKaggle2019 with or without noise filtering (NF).}
\label{tb:nf_on_2019}
\begin{tabular}{cccc}
\toprule[1pt]
           & $M^1$(Lomgel) & $M^2$(CQT) & $M^1 + M^2$ \\
\midrule           
Without NF & 0.6935\tiny{$\pm$0.0023}   & 0.6815\tiny{$\pm$0.0051} & 0.7025\tiny{$\pm$0.0030} \\
NF         & \textbf{0.7105}\tiny{$\pm$0.0055}   & \textbf{0.7060}\tiny{$\pm$0.0044}  & \textbf{0.7195}\tiny{$\pm$0.0027} \\
\bottomrule[1pt]
\end{tabular}
\end{table}

\subsection{Choice of Cross-Representation}
\label{sub:exp_cr}

We explore the performances of different representations and their combinations. Firstly, we inspect the performances of the four most commonly used representations: Spec, MFCC, Logmel and CQT. 
For each kind of representations, we train two same networks (for example, $M^1$) with different random initialization and report their ensemble results. All ensemble method in our experiments is simply add the output probability. Further, we investigate the complementarity of different combinations. We average the output probability of two peer networks trained with different representations. Higher performance of cross-representation is more beneficial for the subsequent data selection process. Here, all the experiments do not use noise filtering component.

Table \ref{tb:exp_cr} shows the accuracy and mAP@3 for different representation on the FSDKaggle2018 dataset. For the single representation, Logmel performs best among the four kinds of representations we have tried, followed by CQT. As Logmel, MFCC and CQT highlight the representation in the low-frequency area compared with linear power spectrogram, it is helpful for improving classification performance. While MFCC further compresses the feature by the discrete cosine transform, this may cause degradation of result. For the cross-representations, the combination of Logmel and CQT achieved 89.11\% accuracy and 92.43\% mAP@3, which is the best combination. The two networks could learn more complementary information from Logmel and CQT so that introduce more divergence for two networks. Therefore, we employ Logmel and CQT as $Rep^1$ and $Rep^2$ in CrossFilter as cross representations.

\begin{table}[t]
\centering
\setlength{\tabcolsep}{10pt}
\caption{Comparison of dataset filtering method and CrossFilter.}
\label{tb:data_filtering}
\begin{tabular}{cccc}
\toprule[1pt]
           & FSD-2018 & FSD-2019  \\
\midrule
Dataset Filtering & 95.07\tiny{$\pm$0.40} & 0.7064\tiny{$\pm$0.0046}   \\
CrossFilter       & \textbf{95.59}\tiny{$\pm$0.20} & \textbf{0.7195}\tiny{$\pm$0.0027}     \\
\bottomrule[1pt]
\end{tabular}
\end{table}

\begin{table}[t]
\centering
\setlength{\tabcolsep}{3pt}
\caption{Comparison of accuracy and mAP@3 without (w/o) or with (w/) using multi-task learning on FSDKaggle2018. Also, different criterions for $loss_c$ and $loss_n$ have been compared.}
\label{tb:mtl}
\begin{tabular}{c|ccccc}
\toprule[1pt]
\multirow{2}{*}{} & & \multicolumn{2}{c}{Logmel} & \multicolumn{2}{c}{CQT} \\
                  & & ACC         & mAP@3       & ACC        & mAP@3      \\
\midrule

\multirow{3}{*}{w/o MTL}
& $CCE$  & 89.15\tiny{$\pm$0.32}    & 92.32\tiny{$\pm$0.38}  & 88.73\tiny{$\pm$0.38}  & 92.07\tiny{$\pm$0.38} \\
& $MAE$  & 82.18\tiny{$\pm$0.64}    & 85.50\tiny{$\pm$0.78}  & 79.49\tiny{$\pm$0.65}  & 84.56\tiny{$\pm$0.60} \\
& $L_q$  & 84.34\tiny{$\pm$0.58}    & 89.72\tiny{$\pm$0.43}  & 83.08\tiny{$\pm$0.49}  & 88.23\tiny{$\pm$0.22} \\
\midrule
\multirow{3}{*}{w/ MTL}
&$loss_n = CCE$          & 89.63\tiny{$\pm$0.14}    & 92.71\tiny{$\pm$0.21}  & 89.38\tiny{$\pm$0.40}  & 92.55\tiny{$\pm$0.23}  \\

&$loss_n = MAE$         & 88.50\tiny{$\pm$0.60}     & 91.63\tiny{$\pm$0.63}   & 87.42\tiny{$\pm$0.50}   & 90.93\tiny{$\pm$0.24}  \\
&$loss_n = L_q$          & \textbf{90.85}\tiny{$\pm$0.12}     & \textbf{93.04}\tiny{$\pm$0.40}   & \textbf{90.67}\tiny{$\pm$0.41}   & \textbf{92.87}\tiny{$\pm$0.54}  \\
\bottomrule[1pt]
\end{tabular}
\end{table}

\begin{figure*}[t]
\centering
\setlength{\tabcolsep}{1pt}
\setlength\extrarowheight{3pt}
\includegraphics[width=2.05\columnwidth]{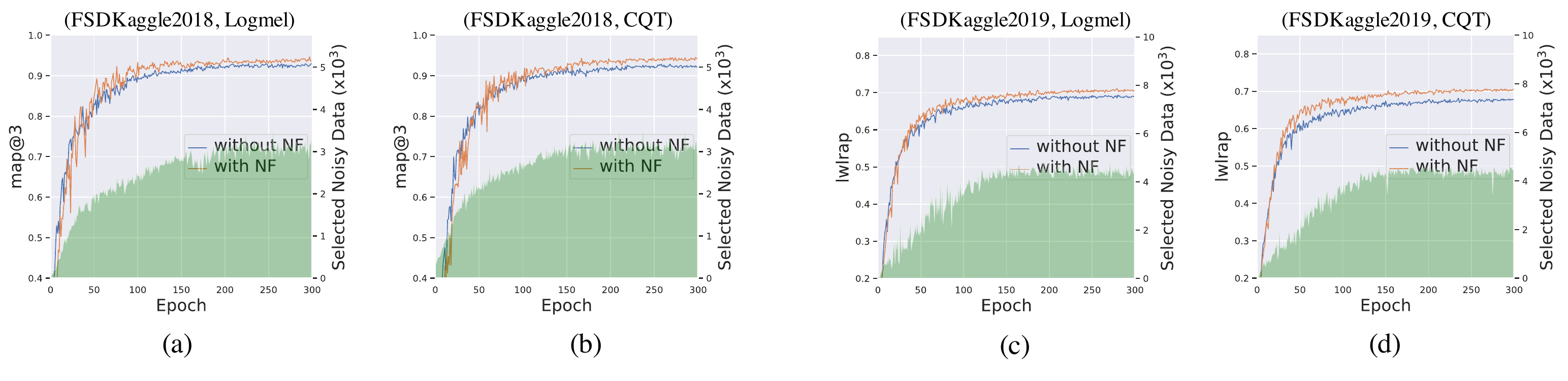}
\caption{(a), (b) Results on FSDKaggle2018: mAP@3 against epoch with or without Noise Filtering. (a) $M^1$ with Logmel, (b) $M^2$ with CQT.
(c), (d) Results on FSDKaggle2019: lwlrap score against epoch with or without NF. (c) $M^1$ with Logmel, (d) $M^2$ with CQT. The green parts at the bottom of the figures show the number of pseudo curated data selected by another network from $\mathcal{N}$.}
\label{fig2}
\end{figure*}

\subsection{Effect of Noise Filtering}

Table~\ref{tb:nf_on_2018} and~\ref{tb:nf_on_2019} show the improvements by Noise Filtering (NF). The settings in the experiments keep the same except Noise Filtering (NF). 
Network $M^1$ is trained with Logmel representation, and $M^2$ is trained with CQT representation. For comparison, we evaluate the performances of two networks ($M^1$ and $M^2$) and their ensamble ($M^1 + M^2$) without using the NF.
On the FSDKaggle2018 dataset, with NF applied, the mAP@3 scores get obvious improvement on both $M^1$ and $M^2$. Noise filtering improves the integral system performance by 1.51\%. Similarly, on the FSDKaggle2019 dataset, NF improves the lwlrap score for both single networks. The couple networks achieve 0.017 improvement due to Noise Filtering.

Figure~\ref{fig2} shows the performance curve during training with NF used or not. It can be seen that the performances with NF get obvious improvement as epoch increasing. The results are conducted on the test set of the two experimental datasets. The green parts at the bottom of the figures show the number of pseudo curated data selected by peer network from $\mathcal{N}$. It can be seen that the number of instances moved from $\mathcal{N}$ to $\mathcal{C}^r$ increase at the beginning of training process and fall to a fixed range when the network gradually converges. For the FSDKaggle2018 dataset, we could select about 3k noisy data, accounting for about half of all noisy labeled data. For the FSDKaggle2019 dataset, the selected noisy data is about 4k, accounting for 40\% of all noisy labeled data. On the one hand, this indicates that the distribution of the noisy subset in FSDKaggle2018 is more consistent with the curated subset than FSDKaggle2019.

After the learning process of CrossFilter, we will obtain new bigger curated subsets $\mathcal{C}^1$ and $\mathcal{C}^2$, where the labels in them are less noisy compared with the whole dataset. From this aspect, our NF method is employable as a dataset filtering method. We inspect the results training on the final curated subsets and compare with the CrossFilter. We use the same network architecture and hyper-parameter to train two peer networks with $\mathcal{C}^1$ and $\mathcal{C}^2$ respectively and report the ensemble results in Table \ref{tb:data_filtering}. Results of dataset filtering method degrade on both experimental datasets. We suppose that our method could be regarded as a form of curriculum learning \cite{bengio2009curriculum,jiang2015self}. It learns from more reliable data at first and gradually increases the number of less reliable data. If we use the whole $\mathcal{C}^r$ from the beginning, parameters of neural networks may get stuck in a bad local optimum.

\subsection{Effect of Multi-Task Learning}
\label{sec:exp_MTL}

To observe the effect of multi-task learning, we perform experiments with a peer network of our framework. On FSDKaggle2018 dataset, the effectiveness of MTL is demonstrated with both Logmel and CQT representations in Table~\ref{tb:mtl}.
As we only discuss the effects of multi-task learning component here, the NF is fairly not used in all experiments.
First, we inspect the performance when not using MTL. All the data, including curated subset and noisy subset, are used to train the network with only one classifier. The accuracy and mAP@3 with different $loss_c$ are reported. It is shown that CCE loss performs better than MAE and $L_q$ losses. For Logmel and CQT, we get 92.32\% and 92.07\% mAP@3 respectively without MTL. 
Next, MTL is employed with different $loss_n$ functions. The $loss_n$ is chosen as CCE, MAE and $L_q$ loss respectively and $loss_c$ remains CCE. For the $L_q$ loss, hyper-parameter $q$ is set as 0.5 because it is more appropriate for noisy data according to \cite{fonseca2019learning}. We can see that, the combination of CCE and $L_q$ under multi-task learning has achieved the better results than using CCE or $L_q$ alone, getting 93.04\% and 92.87\% mAP@3 for Logmel and CQT representations respectively. Compared with the results of without using MTL, MTL generally brings improvement, even if CCE applied on both $loss_c$ and $loss_n$. This implies that it is beneficial to decompose the classification on noisy subset $\mathcal{N}$ to curated subset $\mathcal{C}$ to different tasks for optimization. 

\begin{figure}[t]
\centering
\includegraphics[width=0.95\columnwidth]{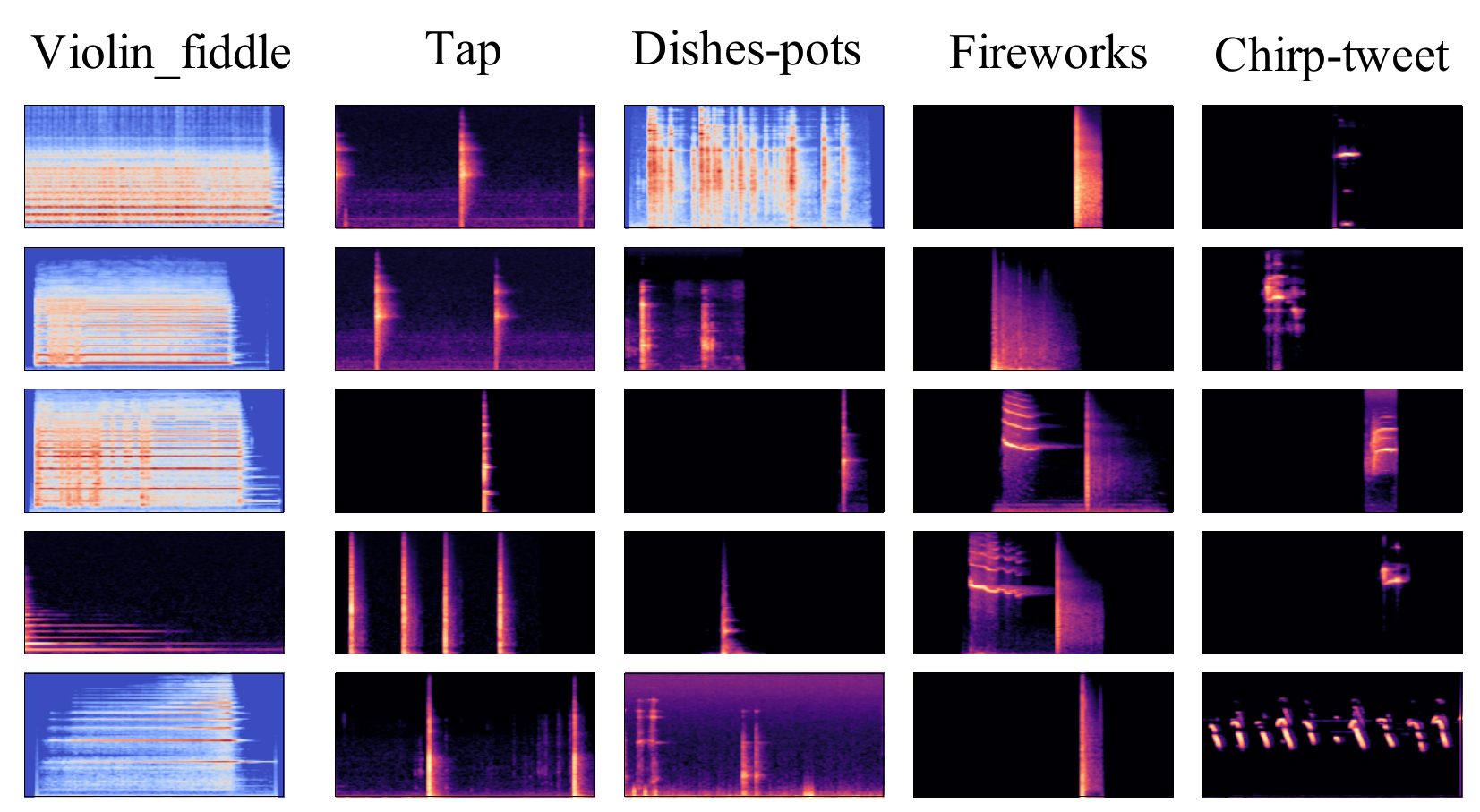} 
\caption{Long-lasting and short-lasting audio clips.}
\label{fig:long_short_lasting}
\end{figure}


\section{Discussion and Conclusion}
\label{sec:conclusion}
Our proposed approach effectively improves the audio tagging performance with noisy labels. While in our experiments, we find the performance is not very satisfactory in two cases and we perform error analysis here. On one hand, the transient or short-lasting audio instance is still hard to be recognized. In our surroundings, many sound events are short-lasting, they usually stay for a short time, for example: tap, collision sound of dishes and pots, firework and chirp. Fig. \ref{fig:long_short_lasting} shows the Logmel representations of these categories. 
Though our method achieves better results in these categories than previous works, the performances on these categories are much lower compared with other long-lasting sounds (e.g. Violin). Due to the blank periods in audio clips, the total time of these categories is shorter than others. Although our method could do sample selection in a class-balanced way, we do not guarantee that the total time of each class is the same. Unbalance audio time may cause poor performance for the neural networks.
On the other hand, the fine-grained classes confusion causes some incorrect tagging. The sound categories are specified as a hierarchical graph. For example, guitar, sitar and ukulele are all belong to musical instruments category but they have their own independent labels. In our system, among the instances of \textit{Female-Speech}, 5.3\% instances are misidentified as \textit{Male-Speech} and 6.2\% instances are misidentified as \textit{Child-Speech}. And 12.6\% of the \textit{Child-Speech} are misjudged as \textit{Crowd}. How to better distinguish these fine-grained audio categories might be a potential research issue in the future.

In conclusion, audio tagging under noisy labels is still a challenging task. In this paper, we propose a novel LNL framework for the audio tagging task. It increases the system robustness under noisy labels with three components: cross-representation, noise filtering, and multi-task learning. CrossFilter employs two kinds of audio representations as input. Meanwhile, with the cooperative learning of two peer neural networks, more reliable data are picked into the curated subset and less reliable data are left in the noisy subset. Then we use the multi-task learning on curated and noisy subsets with different loss functions. In our experiments, we show the efficacy of the framework on various audio tagging datasets. Ablation studies are conducted to demonstrate the effectiveness of each component of the framework.

\appendices


\section*{Acknowledgment}

This work is supported by the National Grand R\&D Plan of China (Grant No. 2016YFB1000101), the General Program of National Natural Science Foundation of China (81973244) and the National Defense Science and Technology Innovation Special Zone Project.

\ifCLASSOPTIONcaptionsoff
  \newpage
\fi



\bibliographystyle{IEEEtran}
\bibliography{IEEE_ref}




%






\end{document}